\documentclass[aps,prb,twocolumn,showpacs,groupedaddress]{revtex4}%
\usepackage{amsmath}
\usepackage{graphics}
\usepackage{graphicx}
\usepackage{epsfig}
\usepackage{amssymb}
\usepackage{amsfonts}%
\providecommand{\U}[1]{\protect\rule{.1in}{.1in}}

\begin{document}
\title{Pressure induced FFLO instability in multi-band superconductors}
\author{Igor T. \surname{Padilha} and Mucio A. \surname{Continentino}}
\affiliation{Instituto de F\'{\i}sica, Universidade Federal
Fluminense} \affiliation{Campus da Praia Vermelha, 24210-340,
Niter\'oi, RJ, Brazil} \email{mucio@if.uff.br}

\begin{abstract}
Multi-band systems as intermetallic and heavy fermion compounds have
quasi-particles arising from different orbitals at their Fermi surface. Since
these quasi-particles have different masses or densities, there is a natural
mismatch of the Fermi wave-vectors associated with different orbitals. This
makes these materials potential candidates to observe exotic superconducting
phases as Sarma or FFLO phases, even in the absence of an external magnetic
field. The distinct orbitals coexisting at the Fermi surface are generally
hybridized and their degree of mixing can be controlled by external pressure.
In this Communication we investigate the existence of an FFLO phase in a
two-band BCS superconductor controlled by hybridization. At zero temperature,
as hybridization (pressure) increases we find that the BCS state becomes
unstable with respect to an inhomogeneous superconducting state characterized
by a single wave-vector $q$.

\end{abstract}

\maketitle

Asymmetric superfluidity refers to Cooper pairing in systems with mismatched
Fermi surfaces. This phenomenon comprises the FFLO \cite{fflo} type of ground
states where the mismatch between bands with different spin orientations is
produced by an external magnetic field. It also occurs in cold atom systems
where the mismatch is due to a different numbers of interacting fermions
\cite{nature,tsai}. Besides, it may appear in the interior of neutron stars
where the pairing of up and down quarks in different numbers can give rise to
color superconductivity (see for example, Refs. \cite{casalbuoni,livro}).

In multi-band metallic systems as inter-metallic compounds and heavy fermions,
electrons arising from distinct atomic orbitals coexist at a common Fermi
surface \cite{mgb2,bianchi}. Since these electrons have different effective
masses or occur in different numbers per atom, there is a natural mismatch of
the Fermi wave-vectors of these quasi-particles. As a consequence, we may
expect to find the physics associated with asymmetric superconductivity in
these systems even in the absence of an external magnetic field. In these
materials, the wave functions of electrons in different orbitals hybridize due
to their overlap. In particular, the mismatch of the Fermi wave-vectors is
affected by hybridization. Since pressure controls hybridization
\cite{japiassu}, we show that in multi-band superconductors it plays a role
similar to that of an external magnetic field in the study of FFLO phases. It
has the advantage that this pressure induced FFLO phase does not compete with
orbital effects which arise when applying an external magnetic field to a superconductor.

The problem of superconductivity in systems with overlapping bands
was treated originally by Suhl, Matthias and Walker \cite{SMW}.
However, these authors did not consider inter-band pairing as this
is negligible in the case where the critical temperature is much
smaller than the effective inter-band splitting.

Recently, we have investigated asymmetric superconductivity in
multi-band metallic systems in the presence of intra and inter-band
interactions \cite{nosso}. We have studied the different types of
homogeneous ground states which appear as hybridization is changed.
In the inter-band case, as hybridization increases there is a first
order transition from the BCS state \cite{bcs} to the normal state.
In between these states there is a gapless metastable phase with
similarities to the Sarma phase \cite{sarma} which has had renewed
interest in recent years \cite{caldas,livro}. The instability of the
BCS state is related to the appearance of a soft mode at a
characteristic wave-vector \cite{nosso,caldas}. This suggests that
an alternative ground state as hybridization increases is an
inhomogeneous superconductor of the FFLO type. In this paper we
investigate the existence of such state. Differently from Ref.
\cite{SMW}, we consider the situation where the dispersion relations
of the bands overlap at the Fermi surface, such that, their Fermi
wave-vectors are equal. In this case inter-band interactions must be
taken into account.

The effective Hamiltonian describing the two-bands metallic system,
hybridization and pairing of quasi-particles with a net momentum $q$ is given
by,
\begin{align}
\mathcal{H}_{eff}  &  =\sum_{k}\left(  \epsilon_{k}^{a}a_{k}^{+}a_{k}%
+\epsilon_{k}^{b}b_{k}^{+}b_{k}\right) \nonumber\\
&  +\sum_{k}\left(  \Delta_{q}a_{k+\frac{q}{2}}^{+}b_{-k+\frac{q}{2}}%
^{+}+\Delta_{q}^{\ast}b_{-k+\frac{q}{2}}a_{k+\frac{q}{2}}\right) \label{1}\\
&  +\sum_{k}V_{k}\left(  a_{k+\frac{q}{2}}^{+}b_{k+\frac{q}{2}}+b_{k+\frac
{q}{2}}^{+}a_{k+\frac{q}{2}}\right) \nonumber
\end{align}
where the inhomogeneous superconducting order parameter is,
\begin{equation}
\Delta_{q}=-g\sum\left\langle b_{-k+\frac{q}{2}}a_{k+\frac{q}{2}}\right\rangle
. \label{1.1}%
\end{equation}
with $g$ the strength of the attractive interaction. The dispersion of the
quasi-particles is given by,
\begin{equation}
\epsilon_{k}^{i}=\xi_{i}\left(  k\right)  -\mu_{i},i=a,b \label{1.2}%
\end{equation}
where,
\begin{equation}
\xi_{i}\left(  k\right)  =\alpha_{i}k^{2},\left\{
\begin{array}
[c]{c}%
\alpha_{a}=1\\
\alpha_{b}=\alpha=\frac{m_{a}}{m_{b}}%
\end{array}
\right.  \label{1.3}%
\end{equation}
and $\alpha<1$ is the ratio of the effective masses.

The Green's function method is used to obtain the BCS-like order parameter
\begin{equation}
\left\langle b_{-k+\frac{q}{2}}a_{k+\frac{q}{2}}\right\rangle \!=\!\int\!
d\omega f(\omega)\left[  \operatorname{Im}\left\langle \left\langle
a_{k+\frac{q}{2}};b_{-k+\frac{q}{2}}\right\rangle \right\rangle _{\omega}
\right] . \label{2}%
\end{equation}
where $f(\omega)$ is the Fermi function.

In order to calculate the relevant Greens functions we obtain their equations
of motion. In particular for the anomalous Greens function $\left\langle
\left\langle a_{k+\frac{q}{2}};b_{-k+\frac{q}{2}}\right\rangle \right\rangle
_{\omega} $ this is given by,
\begin{align}
&  \omega\left\langle \left\langle a_{k+\frac{q}{2}};b_{-k+\frac{q}{2}%
}\right\rangle \right\rangle _{\omega} =\left\langle \left\langle \left[
a_{k+\frac{q}{2}},\mathcal{H}_{eff}\right]  ;b_{-k+\frac{q}{2}}\right\rangle
\right\rangle _{\omega} +\nonumber\\
&  \frac{1}{2\pi}\left\langle \left\{  a_{k+\frac{q}{2}},b_{-k+\frac{q}{2}%
}\right\}  \right\rangle . \label{3}%
\end{align}

After some long calculations we obtain for the anomalous Green's function,
\begin{equation}
\left\langle \left\langle a_{k+\frac{q}{2}};b_{-k+\frac{q}{2}}\right\rangle
\right\rangle _{\omega} =\frac{D_{x}(\omega)}{D(\omega)} \label{4}%
\end{equation}
with
\begin{equation}
D_{x}(\omega)=\Delta_{q}\!\left[  \left(  \omega-\epsilon_{k-\frac{q}{2}}%
^{b}\right) \!\! \left(  \omega+\epsilon_{-k+\frac{q}{2}}^{a}\right)
\!-\!\left(  \left\vert \Delta_{q}\right\vert ^{2}\!-V_{k}^{2}\right)
\right]  \label{5}%
\end{equation}
and
\begin{align}
&  D(\omega)\!=\!\left(  \omega\!+\!\epsilon_{\!-\!k\!+\!\frac{q}{2}}%
^{b}\right)  \!\! \left(  \omega\!-\!\epsilon_{k\!-\!\frac{q}{2}}^{a}\right)
\!\! \left(  \omega\!-\!\epsilon_{k\!+\!\frac{q}{2}}^{b}\right)  \!\! \left(
\omega\!+\!\epsilon_{\!-\!k\!+\!\frac{q}{2}}^{a}\right) \nonumber\\
&  \!-\!\!V_{k}^{2}\!\left[  \!\left(  \omega\!+\!\epsilon_{\!-\!k\!+\!\frac
{q}{2}}^{b}\right)  \!\! \left(  \omega\!+\!\epsilon_{\!-\!k\!+\!\frac{q}{2}%
}^{a}\right)  \!\!+\!\!\left(  \omega\!-\!\epsilon_{k\!-\!\frac{q}{2}}%
^{a}\right)  \!\! \left(  \omega\!-\!\epsilon_{k\!+\!\frac{q}{2}}^{b}\right)
\! \right] \nonumber\\
&  \!-\!\left\vert \Delta_{q}\right\vert ^{2}\!\!\left[  \!\left(
\omega\!-\!\epsilon_{k\!+\!\frac{q}{2}}^{b}\right)  \!\! \left(
\omega\!+\!\epsilon_{\!-\!k\!+\!\frac{q}{2}}^{a}\right)  \!+\!\left(
\omega\!+\!\epsilon_{\!-\!k\!+\!\frac{q}{2}}^{b}\right)  \!\! \left(
\omega\!-\!\epsilon_{\!k\!+\!\frac{q}{2}}^{a}\right)  \!\right] \nonumber\\
&  \!+\!\left(  V_{k}^{2}\!-\!\left\vert \Delta_{q}\right\vert ^{2}\right)
^{2}. \label{6}%
\end{align}
The poles of the Green's function, $D(\omega)\!=\!0$, in Eq. \ref{6} yield the
excitations of the system. Substituting the dispersion relation of the bands,
\begin{align}
\epsilon_{\pm k+\frac{q}{2}}^{a}  &  =k^{2}+\frac{q^{2}}{4}\pm\overrightarrow
{k}.\overrightarrow{q}-\mu_{a}\nonumber\\
\epsilon_{\pm k+\frac{q}{2}}^{b}  &  =\alpha k^{2}+\alpha\frac{q^{2}}{4}%
\pm\alpha\overrightarrow{k}.\overrightarrow{q}-\mu_{b}\nonumber
\end{align}
in Eq.\ref{6}, we obtain a complete fourth degree equation for the energy
$\omega$ of the excitations,
\begin{equation}
D=\omega^{4}+b\omega^{3}+c\omega^{2}+d\omega+e=0 \label{8}%
\end{equation}
where
\begin{align}
b  &  =-2v_{F}qX\left(  1+\alpha\right) \label{blah}\\
c  &  =-\left[  \epsilon_{k}^{a2}+\epsilon_{k}^{b2}+2\left(  V_{k}%
^{2}+\left\vert \Delta_{q}\right\vert ^{2}\right)  \right] \nonumber\\
d  &  =2v_{F}qX\left[  \epsilon_{k}^{b2}+\alpha\epsilon_{k}^{a2}+\left(
1+\alpha\right)  \left(  V_{k}^{2}+\left\vert \Delta_{q}\right\vert
^{2}\right)  \right] \nonumber\\
e  &  =\left[  \epsilon_{k}^{a}\epsilon_{k}^{b}-\left(  V_{k}^{2}-\left\vert
\Delta_{q}\right\vert ^{2}\right)  \right]  ^{2}\nonumber\\
X  &  =\frac{\overrightarrow{k}.\overrightarrow{q}}{kq}=\cos\theta\nonumber
\end{align}
with $v_{F}$ the Fermi velocity and we have neglected terms of $O(q^{2})$ as usual.

In order to solve this equation we introduce the change of variable,
\begin{equation}
\omega\rightarrow u-\frac{b}{4}=u+v_{F}q\frac{\left(  1+\alpha\right)  }%
{2}\cos\theta\label{9}%
\end{equation}
which yields a \textit{depressed} equation of the fourth degree
\begin{equation}
u^{4}+\beta u^{2}+\gamma u+\lambda=0 \label{depressed}%
\end{equation}
where
\begin{align}
\beta &  =\frac{-3b^{2}}{8}+c=-2(V^{2}+\Delta_{q}^{2})-\epsilon_{k}%
^{a2}-\epsilon_{k}^{b2}\nonumber\\
\gamma &  =\frac{b^{3}}{8}-\frac{bc}{2}+d=-qv_{F}X(1-\alpha)(\epsilon_{k}%
^{a2}-\epsilon_{k}^{b2})\nonumber\\
\lambda &  =\frac{-3b^{4}}{256}+\frac{cb^{2}}{16}-\frac{bd}{4}+f=(\epsilon
_{k}^{a}\epsilon_{k}^{b}-V^{2}+\Delta_{q}^{2})^{2}\nonumber
\end{align}
up to linear terms in $q$. In the case $V=0$, $\alpha=1$, $\epsilon_{k}%
^{a}=\epsilon_{k}^{b}$, the fourth order equation reduces to a product of two
identical second order equations. The roots of this second order equation
yield the excitations found in the usual FFLO problem.

The problem above is still quite intractable. This is due to the different
masses ($\alpha\neq1$) of the quasi-particles that in combination with mixing
has a very strong destabilizing effect on the FFLO state. The effects of
hybridization are stronger at the points in $k$-space where the bands cross,
i.e., for $\epsilon_{k_{c}}^{a}=\epsilon_{k_{c}}^{b}$. Analytical progress can
be done if we assume the case of homotectic bands, i.e., we take $\epsilon
_{k}^{b}=\alpha\epsilon_{k}^{a}$, and $\epsilon_{k}^{a}=\epsilon_{k}$. The
crossing of the bands takes place exactly at the Fermi surface, at
$\epsilon_{k}^{i}=0$. Furthermore, to make analytical progress we consider
that the ratio between the masses of the quasi-particles $\alpha$ is very
close to unity, i.e., we write $\alpha=1-\varepsilon$, and neglect terms of
order $\epsilon^{2}$. In this case we can find a solution for the depressed
fourth order equation given by Eq. \ref{depressed}.

The energies of the excitations in this case are given by $\omega=\omega
_{12}^{\pm}(k)$, where,
\begin{equation}
\omega_{12}^{\pm}(k)=\pm\omega_{12}+\delta\mu\label{poles}%
\end{equation}
with,
\begin{equation}
\omega_{12}(k)=\sqrt{A_{k}\pm\sqrt{B_{k}}.}%
\end{equation}
The quantity $\delta\mu=-b/4=v_{F}q[\left(  1+\alpha\right)  /2]\cos\theta$.
Also,
\begin{equation}
A_{k}=(1-\varepsilon)\epsilon_{k}^{2}+V^{2}+\Delta_{q}^{2}+O[\epsilon]^{2}
\label{ak}%
\end{equation}
and
\begin{equation}
B_{k}=4V^{2}[(1-\varepsilon)\epsilon_{k}^{2}+\Delta_{q}^{2}]+O[\epsilon]^{2}.
\label{bk}%
\end{equation}
These equations yield
\begin{equation}
\omega_{12}(k)=\xi_{k}\pm V \label{dispersion}%
\end{equation}
where
\[
\xi_{k}=\sqrt{(1-\varepsilon)\epsilon_{k}^{2}+\Delta_{q}^{2}}.
\]
When calculating the gap function $\Delta_{q}$ we find, after a change of
variables, the following integral,
\[
G_{k}(\delta\mu)=\frac{1}{2\pi}\int_{-\infty}^{+\infty}d\omega D_{x}%
(\omega+\delta\mu)\operatorname{Im}\left[  \frac{1}{D(\omega)}\right]
f(\omega+\delta\mu)
\]
where $f(\omega)$ is the Fermi function, $D_{x}(\omega)$ is given by Eq.
\ref{5} above and the denominator of the anomalous Greens function is given
by:
\[
D=(\omega^{2}-\omega_{1}^{2})(\omega^{2}-\omega_{2}^{2}).
\]
Using that,
\begin{align*}
\frac{1}{(\omega^{2}-\omega_{1}^{2})(\omega^{2}-\omega_{2}^{2})}  &  =\frac
{1}{8V\xi_{k}}\{\frac{1}{\omega_{1}}\left[  \frac{1}{\omega-\omega_{1}}%
-\frac{1}{\omega+\omega_{1}}\right]  -\\
&  \frac{1}{\omega_{2}}\left[  \frac{1}{\omega-\omega_{2}}-\frac{1}%
{\omega+\omega_{2}}\right]  \}
\end{align*}
Recalling that in the equation above, $\omega\rightarrow\omega+i\epsilon$, and
taking the imaginary part, we obtain that $G_{k}(\delta\mu)$ is a sum of three
terms, $G_{k}(\delta\mu)=G_{k}^{1}(\delta\mu)+G_{k}^{2}(\delta\mu)+G_{k}%
^{3}(\delta\mu)$ with,
\[
G_{k}^{1}(\delta\mu)=\frac{\Delta_{q}}{4\xi_{k}}\left\{  2-\sum_{\sigma
}[f(E_{k\sigma}^{1})+f(E_{k\sigma}^{2})]\right\}  ,
\]
\begin{align*}
G_{k}^{2}(\delta\mu)  &  =\frac{-\Delta_{q}[(1-\alpha)\epsilon_{k}%
+2\alpha\overrightarrow{k}.\overrightarrow{q}]}{8V\xi_{k}}\times\\
&  \left\{  \sum_{j=1,2}(-1)^{j-1}[f(E_{k+}^{j})+f(E_{k-}^{j})]\right\}
\end{align*}
and
\begin{align*}
G_{k}^{3}(\delta\mu)  &  =\frac{\Delta_{q}(1+\alpha)\epsilon_{k}}{8\xi_{k}%
(\xi_{k}^{2}-V^{2})}\overrightarrow{k}.\overrightarrow{q}\times\\
&  \left\{  2+\sum_{j=1,2}(-1)^{j-1}[f(E_{k+}^{j})-f(E_{k-}^{j})]\right\}  .
\end{align*}
where $E_{k\sigma}^{1}=\xi_{k}+\sigma(V+\delta\mu)$ and $E_{k\sigma}^{2}%
=\xi_{k}+\sigma(V-\delta\mu)$ with $\sigma=\pm$. We have omitted terms of
$O(q)^{2}$ and $O(\varepsilon)^{2}$. When calculating the gap equation
$\Delta_{q}=\sum_{\overrightarrow{k}}G_{k}(\delta\mu)$ at zero temperature,
the Fermi functions are expressed in terms of $\theta$ functions and this
imposes severe restrictions on the sums over $\overrightarrow{k}$. When these
sums are performed and angular integrations are carried out, the only
contribution which remains is that arising from $G_{k}^{1}(\delta\mu)$. The
gap equation can finally be written as,
\begin{equation}
-1+\frac{g}{2}\!\int\!\!\!\frac{d^{3}k}{(2\pi)^{3}}\frac{1}{\xi_{k}}=\frac
{g}{4}\! \int\!\!\!\frac{d^{3}k}{(2\pi)^{3}}\frac{1}{\xi_{k}}\sum_{\sigma
}[\theta(-E_{k\sigma}^{1})+\theta(-E_{k\sigma}^{2})] \label{gap}%
\end{equation}
Subtracting the $T=0$ gap equation for a BCS superconductor,
\begin{equation}
-1+\frac{g}{2}\int\frac{d^{3}k}{(2\pi)^{3}}\frac{1}{\sqrt{\alpha\epsilon
_{k}^{2}+\Delta_{0}^{2}}}=0 \label{gap0}%
\end{equation}
with $\alpha\approx1$, from the left hand side of Eq. \ref{gap}, we obtain in
the weak coupling approximation,
\[
\frac{g\rho}{2\sqrt{\alpha}}\ln\frac{\Delta_{0}}{\Delta_{q}}=\frac{g}{4}%
\int\frac{d^{3}k}{(2\pi)^{3}}\frac{1}{\xi_{k}}\sum_{\sigma}[\theta
(-E_{k\sigma}^{1})+\theta(-E_{k\sigma}^{2})]
\]
where $\rho$ is the density of states at the Fermi level. The integrals over
$k$ ($\int dk$) on the right hand side are performed taking into account the
constraints imposed by the $\theta$ functions. They yield,
\[
\frac{g\rho}{4\sqrt{\alpha}}\sum_{\sigma}\int\frac{d\Omega}{4\pi}\sinh
^{-1}\left[  \frac{\sqrt{(V+\sigma\delta\mu)^{2}-\Delta_{q}^{2}}}{\Delta_{q}%
}\right] .
\]
This equation has real solutions only if $V+v_{F}^{\ast}q>\Delta_{q}$ where
$v_{F}^{\ast}=v_{F}\left(  1+\alpha\right)  /2$. Let us consider the case
$\sigma=-1$, Recalling that $\delta\mu=v_{F}^{\ast}q\cos\theta$, the integral
above can be rewritten as,
\[
\frac{g\rho}{4\sqrt{\alpha}}\frac{1}{2v_{F}^{\ast}q}\int_{-v_{F}^{\ast}%
q}^{v_{F}^{\ast}q}dx\sinh^{-1}\left[  \frac{\sqrt{(V+x)^{2}-\Delta_{q}^{2}}%
}{\Delta_{q}}\right]
\]
where we used the change of variables, $x=-v_{F}^{\ast}q\cos\theta$. In fact
the integrals are independent of $\sigma$ and the result is simply twice that
for a given sign. Respecting the limits of integration in different cases to
obtain a real result, the final gap equation is given by,
\begin{equation}
\frac{g\rho}{2\sqrt{\alpha}}\ln\frac{\Delta_{0}}{\Delta_{q}}=\frac{g\rho
}{4\sqrt{\alpha}}\frac{\Delta_{q}}{v_{F}^{\ast}q}\left[  G(\frac{v_{F}^{\ast
}q+V}{\Delta_{q}})+G(\frac{v_{F}^{\ast}q-V}{\Delta_{q}})\right]  \label{final}%
\end{equation}
where $G(x)$ is the function \cite{izuyama},
\begin{align*}
G(x)  &  =x\cosh^{-1}x-\sqrt{x^{2}-1},|x|>1\\
&  =0,|x|\leq1\\
&  =-G(-x),x<0.
\end{align*}
Notice that the mass ratio $\alpha$ cancels out explicitly in the gap
equation, Eq. \ref{final}. It's role at least for $\alpha\thickapprox1$ is
just to renormalize the Fermi velocity. From this equation we find that for
the FFLO state to be a solution it is necessary that $\overline{q}%
=q/(V/v_{F}^{\ast})>1$. Also, since $G(|x|\leq1)=0$, the solution for
$V<V_{1}^{c}(\overline{q})=$ $\Delta_{0}/(1+\overline{q})$ is always
$\Delta_{q}=\Delta_{0}$, i.e., the BCS state. Thus a necessary condition for
the FFLO state is $V>V_{1}^{c}(\overline{q})$. The upper critical value of the
hybridization $V_{2}^{c}(\overline{q})$ below which the FFLO state can be a
solution of the gap equation is obtained taking the limit of Eq.\ref{final}
for $\Delta_{q}\rightarrow0$. The results can be expressed as
\cite{casalbuoni,izuyama},
\[
V_{2}^{c}(\overline{q})=\frac{\Delta_{0}e}{2(\overline{q}+1)}\left\vert
\frac{\overline{q}+1}{\overline{q}-1}\right\vert ^{\frac{\overline{q}%
-1}{2\overline{q}}}.
\]
\begin{figure}[th]
\centering{\includegraphics[scale=0.75]{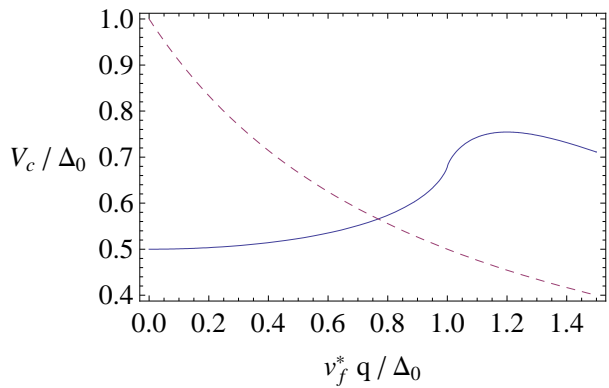}}\caption{{$V_{1}^{c}$}
(dashed) and {$V_{2}^{c}$} as a function of the reduced wave-vector.}%
\label{izuy}%
\end{figure}In Figure \ref{izuy} we plot $V_{1}^{c}(\overline{q})$ and
$V_{2}^{c}(\overline{q})$ as a function of the reduced wave-vector and it is
clear that there is a range of values for the hybridization $V_{c}^{1}%
<V<V_{c}^{2}$ for which a FFLO phase may exist. The maximum value of
$V_{2}^{c}$ occurs for $\overline{q}=\overline{q}_{c}\thickapprox1.2$, which
when substituted in the equation above yields $V_{c}=V_{2}^{c}(\overline
{q}_{c})\thickapprox0.75\Delta_{0}$. This value of $\overline{q}$ is that
which minimizes the free energy in the range of stability of the FFLO phase
\cite{casalbuoni,izuyama}. The value $V_{1}^{c}(\overline{q})$ above marks the
limit of stability of the FFLO phase. The actual value of the hybridization
for which the first order phase transition occurs is obtained considering the
energies of these states. The argument is similar to that of Chandrasekhar and
Clogston \cite{chanclog} to obtain the critical field in BCS superconductors.
Here we have to consider the hybrid bands. In the limit of very small mass
differences their dispersion relations can be easily obtained and are given
by, $\omega_{1,2}=[(1+\alpha)/2]\epsilon_{k}\pm V$. On the other hand the
condensation energy for a system of unequal masses was obtained in
Ref.\cite{yip}. This is similar to that of a system of identical particles
with the mass $m$ replaced by $2m_{r}$, where the reduced mass, $m_{r}%
=m_{a}m_{b}/(m_{a}+m_{b})=m_{a}/(1+\alpha)$ in our notation. The chemical
potential is also modified and given by, $\mu^{\ast}=(\mu_{a}+\mu
_{b})/2=[(1+\alpha)/2]\mu_{a}$. Then the effective particles have dispersion,
$\epsilon_{k}^{\ast}=[(1+\alpha)/2]\epsilon_{k}$. Comparing the condensation
energy of these quasi-particles, $E_{c}=(1/2)\rho^{*} \Delta_{0}^{2}$ with the
energy associated with hybridization, $E_{V}=\rho^{*} V^{2}$, one obtains a
critical hybridization, $V_{c}=\Delta_{0}/\sqrt{2}\thickapprox0.71\Delta_{0}$,
above which BCS superconductivity becomes unstable. In these expressions,
$\rho^{*}$ is the density of states at the Fermi level of particles with
dispersion relation $\epsilon_{k}^{\ast}=[(1+\alpha)/2]\epsilon_{k}$.
Consequently there is a window of values for the hybridization ($0.71\Delta
_{0}<V<0.75\Delta_{0}$) where we can expect a FFLO phase to occur. The
transition at $V_{2}^{c}$ is a continuous second order transition from the
FFLO to the normal state.

Our results have a close similarity to the usual FFLO approach for a
superconductor in an external magnetic field. This was anticipated from the
form of the dispersion relations, Eqs. \ref{dispersion}, where V enters
formally as an external magnetic field. However, the analogy with the usual
FFLO stops there. The Greens functions in the present case have four poles,
instead of two and the numerator of the anomalous Greens function (Eq.
\ref{5}) is much more complex and includes an angular dependence. At the level
of the Hamiltonian, Eq. \ref{1}, $V$ mixes different states and from this
point of view it acts like a \textit{transverse field} and not as a polarizing
longitudinal field. The latter only repopulates the states while the former
changes the nature of the quantum states.

The FFLO phase in condensed matter systems has long been sought.
Here we point out the possibility of attaining an inhomogeneous
superconducting state by applying pressure in a multi-band
superconductor. The existence of quasi-particles belonging to
different orbitals in a common Fermi surface provides a natural
mismatch. It can be controlled by pressure and this, as we have
shown, offers the possibility of finding new inhomogeneous
superconducting states tuning this external parameter.

\acknowledgements{ We would like to thank the Brazilian agencies,
FAPEAM, FAPERJ and CNPq for financial support.The authors thank many
discussions with Dr. Heron Caldas.}


\begin{thebibliography}{99}                                                                                               %
\bibitem {fflo}P. Fulde and R. A. Ferrell, Phys. Rev. \textbf{135}, A550
(1964); A. I. Larkin and Yu N. Ovchinnikov, Sov. Phys. JETP \textbf{20}, 762 (1965).

\bibitem {nature}For a review, see Nature (London) \textbf{416}, 205 (2002).

\bibitem {tsai}L. Mathey, S.-W. Tsai, and A. H. Castro Neto, Phys. Rev. Lett.
97, 030601 (2006).

\bibitem {casalbuoni}R. Casalbuoni and G. Nardulli, Rev. Mod. Phys.,
\textbf{76}, 263 (2004).

\bibitem {livro}see \textit{Pairing in Fermionic Systems} edited by A.
Sedrakian, J. W. Clark and M. Alford, World Scientific, Singapore, 2006.

\bibitem {mgb2}S. Bud'ko, R. H. T. Wilke, M. Angst and P. C. Canfield, Physica
\textbf{C 420}, 83 (2005).

\bibitem {bianchi}A. Bianchi, R. Movshovich, I. Vekhter, P. G. Pagliuso, and
J. L. Sarrao, Phys. Rev. Lett. 91, 257001 (2003).

\bibitem {japiassu}G. M. Japiassu, M. A. Continentino and A. Troper, Phys.
Rev. \textbf{B 45 }, 2986 (1992).

\bibitem{SMW} H. Suhl, B. T. Matthias and L. R. Walker, Phys. Rev.
Lett. \textbf{3}, 552 (1959).

\bibitem {nosso}M. A. Continentino, I. T. Padilha, J. of Phys. Cond. Matter
20, 095216 (2008).

\bibitem {bcs}J. Bardeen, L. N. Cooper and J. R. Schrieffer, Phys. Rev.
\textbf{108}, 1175 (1957).

\bibitem {sarma}G. Sarma, J. Phys. Chem. Solids \textbf{24}, 1029 (1963).

\bibitem {caldas}P. F. Bedaque, H. Caldas and G. Rupak, Phys. Rev. Lett
\textbf{91}, 247002 (2003); H. Caldas, Phys. Rev. \textbf{A 69}, 063602 (2004).

\bibitem {izuyama}S. Takada and T. Izuyama, Prog. Theo. Phys. \textbf{41}, 635 (1969).

\bibitem {chanclog}B. S. Chandrasekhar, Appl. Phys. Lett. \textbf{1}, 7
(1962); A. M. Clogston, Phys. Rev. Lett. \textbf{9}, 266 (1962).

\bibitem {yip}Shin-Tza Wu, C,-H Pao and S. -K. Yip, Phys. Rev. \textbf{B74},
224504 (2006).
\end{thebibliography}
\end{document}